\documentclass[prb,aps]{revtex4}

\usepackage{graphicx}
\usepackage{epsfig}

\newcommand{\ba}{\begin{eqnarray}} 
\newcommand{\ea}{\end{eqnarray}} 
\newcommand{\be}{\begin{equation}} 
\newcommand{\ee}{\end{equation}} 
\newcommand{\bea}{\begin{eqnarray}} 
\newcommand{\eea}{\end{eqnarray}} 

\def\etal{{\it et al}.}

\def\eqnref#1{Eq.(\ref{#1})}
\def\sectref#1{Section \ref{#1}}

\def\figref#1{Figure \ref{#1}}

\begin{document}

\def\CC{{\rm\kern.24em \vrule width.04em height1.46ex depth-.07ex \kern-.30em C}}

\title{Thomas-Fermi Theory of the Hyperfine Constants of Endohedral Fullerene Atoms}

\author{Joshua Schrier and K. Birgitta Whaley}

\affiliation{Department of Chemistry and Pitzer Center for Theoretical
Chemistry, University of California, Berkeley, CA 94720}

\begin{abstract}

We present a modified Thomas-Fermi theory that describes the increase
of the hyperfine coupling constants of endohedrally enclosed atoms.
We use the March boundary conditions corresponding to a positively
charged spherical shell surrounding the nuclear potential to represent
the effect of the fullerene shell.  We obtain quantitative agreement
with experimental data for N@C$_{60}$ and N@C$_{70}$, and find that
fullerene radius dominates over the fullerene charge in its effect on
the hyperfine coupling constants.  We also present predictions for the
hyperfine coupling constants of the endohedral nitrogen fullerenes
between C$_{60}$ and C$_{500}$, and discuss the implications for
proposed quantum computing schemes.

\end{abstract}

\maketitle

\section{Introduction} \label{intro}

The nitrogen endohedral fullerene molecule, N@C$_{60}$, has received a
great deal of
experimental\cite{WWPM98,PWM+98,DHP+99,WGK+00,DKKW00,PWDH02,Dinse02,JDM+03,MTA+04}
and theoretical\cite{LZZ99,Gre00, KND03} study.  Unlike the
metallofullerenes, the endohedral nitrogen acts essentially as a
``free'' atom with a $S=3/2$ quartet ground electronic state, although
spatially restricted by a harmonic-oscillator-like potential within
the fullerene.\cite{PWM+98,WWPM98,Dinse02} Additionally, the fullerene
acts a Faraday cage, shielding the spin of the endohedral nitrogen
atom from stray fields,\cite{DG04} and the sharp ESR spectra indicate
long spin relaxation times,\cite{DHP+99} which have inspired its use
as a potential spin qubit for quantum computation.\cite{PSC01,HMW+02,Harneit02,SL02,Twa03,FT04b}
Experimental\cite{MTA+04} and theoretical\cite{MT04} studies have
indicated that suitable pulse sequences can reduce single qubit errors
in these systems to the order of $10^{-6}$, which is within the
$10^{-4}$ error threshold of Steane.\cite{Steane03}

An interesting physical phenomenon observed for endohedral atom
systems is an increase in the hyperfine coupling constant as compared
to its gas phase value, in particular, an increase of $54.1\%$ for
N@C$_{60}$ and of $49.1\%$ for N@C$_{70}$.\cite{Bucha01} To better
understand the origin of this effect, Kobayashi \etal\,carried out
{\it ab initio} MP2/uc-Huginaga+(2df) calculations, obtaining
enhancements of $77\%$ and $65\%$ for N@C$_{60}$ and N@C$_{70}$,
respectively.\cite{KND03} In general, quantitative calculation of the
hyperfine constant is a difficult problem, even for the free
atom.\cite{Carmichael90} However, a clear qualitative model for the
effects of changes in fullerene size and charge that allows prediction
of these coupling constants would be helpful, not only for
understanding the underlying physics, but also to provide estimates of
potential sources of qubit error in applications to quantum
information processing.  In this paper we introduce a modified
Thomas-Fermi theory for the hyperfine coupling constants in endohedral
fullerenes that is capable of quantitatively reproducing the
experimental results and provides the desired predictive capabilities.
We show that the model is accurate by reproducing the experimentally
observed increase in coupling constants for N@C$_{60}$ and
N@C$_{70}$. We then analyze the independent roles of fullerene cage
size and number of electrons on this enhancement and make predictions
for the hyperfine coupling enhancement for nitrogen atoms endohedral
to fullerenes between C$_{60}$ and C$_{500}$.

\section{Theory}\label{theory}

Following the spherical boundary condition model for highly symmetric molecules developed by
 March,\cite{March52} and later applied to
endofullerenes by Clougherty,\cite{Clougherty96}
we treat the fullerene as a sphere of radius
$r=R$, contributing $n$ $\pi$-electrons and with a charge of $N e$, 
around the nuclear point charge of $Z e$ at the origin.  For example, for
N@C$_{60}$, $Z = 7$ and $n = 60$. Transforming into the reduced
coordinates, $x$, given by $r = b x$ where
\be
b = {{1}\over{4}} {\left ( {{9 \pi^2} \over {2 Z}} \right
)}^{1/3} a_{Bohr}
\ee
and $a_{Bohr}$ is the Bohr radius, we transform the cage radius $R
\to X$.  We then proceed to solve the usual Thomas-Fermi differential
equation\cite{ParrYang}
\be
{{d^{2}\chi} \over {dx^{2}}} = {{\chi}^{3/2} \over {x^{1/2}}}
\ee
with the additional boundary condition,\cite{March52,Clougherty96}
\be
\chi^{\prime}(X^{-}) - \chi^{\prime}(X^{+}) = {Z \over {nX}},
\ee
to account for the spherical shell, in addition to a continuity
condition, $\chi(X^{-}) = \chi(X^{+})$, and the usual boundary
conditions of $\chi(0) = 1$ and $\chi(\infty) = 0$.  Clougherty and
co-workers\cite{Clougherty96,CZ97} have examined the effect of
treating the icosahedral (i.e., non-spherical) nature of the fullerene
by means of a multipole expansion, but we neglect this effect in the
current work.

Next, we use the expression derived by Fermi,\cite{Fermi27,Wong79}
\be\label{Nl} p_{l} = {{4 (2 l + 1)}\over{ 2 \pi \hbar}}
\int\limits_{r_1}^{r_c} \sqrt{ 2 m e V(r) - {\left [ \left (l +
{\frac{1}{2}} \right )\hbar r^{-1} \right ]}^{2} } \, dr \ee which
relates the number of electrons, $p_{l}$, having angular momentum $l$,
to the potential, $V(r) = Ze\chi(r)/r$.  Other expressions for
determining the angular momentum assignments from Thomas-Fermi models
have been developed, but give similar results for low
$Z$.\cite{JL52,Oliph56} The limits of integration are chosen so that
the integrand (and hence the integral) is a positive real number.  In
our case, we take the upper limit of integration to be some tunable
cutoff radius, $r_{c}$, inside the fullerene shell, to count
only the endohedral atom electrons and avoid contributions of the
fullerene electrons to the integral.
We are
primarily interested in the number of unpaired $l=0$ electrons ({\it
vide infra}).  Considering only the ``valence'' $l=0$ electrons, there
is no ambiguity due to Hund's rule, as there is for the $l > 0$
electrons, so we may therefore express the number of fractional, i.e.,
non-integer, number of $l=0$ electrons as 
\be 
p^{frac}_{l=0} =
{\rm frac} \left ( {{{p_{l=0}}\over{2}}}  \right )
\ee
 and the number of unpaired
$l=0$ electrons as 
\be\label{Nunpaired} p^{unpaired}_{l=0} = \left
\lbrace { { p^{frac}_{l=0} \,\,\,\,\,\,\, , \, {\rm
if}\,p^{frac}_{l=0} \leq {\frac{1}{2}}} \atop {1- p^{frac}_{l=0} \, ,
\, {\rm if}\,p^{frac}_{l=0} > {\frac{1}{2}}} } \right. .  \ee

%\noindent Alternatively, one might use a formally spin-polarized Thomas-Fermi theory.\cite{GR88}

The contact-term of the isotropic hyperfine coupling constant,\cite{PBD68,Sadlej} 
\be a =
({\frac{4 \pi}{3}}) g_{e} g_{N} \mu_{B} \mu_{N} \hbar {\langle S_{z}
\rangle}^{-1} \langle \Psi \vert \rho(0) \vert \Psi \rangle, 
\ee 
is
proportional to the electron spin-density (difference between spin up and spin
down densities) at the nucleus, $\rho(0)$
%BW 
(${\langle S_{z} \rangle}^{-1}$). However it is
well known that the Thomas-Fermi method gives an infinite density at
the nucleus, and qualitatively poor results near the nucleus (precluding
extrapolation), so a direct evaluation of $\rho(0)$ will 
fail.\cite{ParrYang} A similar problem occurs in the treatment of the
hyperfine coupling constants by semiempirical quantum chemistry
methods, in which the $2s$ and $3s$ Slater orbital basis functions
erroneously have zero density at the nucleus.\cite{PBD68}  

To avoid this problem, we present a heuristic argument in the language
of atomic orbital theory.  The atom-centered basis functions on a free
atom may each be decomposed into a radial function multiplied by a
spherical harmonic angular function.  For orbitals with $l>0$, a node
occurs at the nucleus, resulting from the properties of the spherical
harmonics, so the electrons in these orbitals have no density at the
nucleus.  This is unchanged when the atom is placed in a spherically
symmetric potential, allowing us restrict our attention to the $l=0$
electrons only.  The magnitude of the spin density at the origin,
$\rho(0)$, is the product of the probability density of finding an
$l=0$ electron at the origin, times the number of unpaired $l=0$ electrons,
$p_{l=0}^{unpaired}$, given by \eqnref{Nunpaired}.  Following the
argument made by Pople \etal,\cite{PBD68} we can consider the basis,
and hence the $l=0$ propability density term, to be
unchanged by the chemical surroundings of the atom, and consider the
changes to occur only in the basis orbital populations.  Similar
arguments have been made in the calculation of host medium effects
(including endohedral fullerene inclusion) on the $L/K$
electron-capture $\beta$-decay ratio of
${}^{7}$Be.\cite{RDS+02,OYM+04} This allows evaluation of the ratio of $\rho(0)$
for the free and endohedral atom cases, as in \eqnref{phi}, below, 
in which
 the unknown probability density term now cancels, leaving the ratio in
terms of $p_{l=0}^{unpaired}$, which we can calculate from
\eqnref{Nunpaired}.

Strictly, the fullerene carbon-atom basis functions may have a small
non-zero value at the nitrogen atom nucleus.  However, if we make a
zero-differential overlap (ZDO) approximation, which is well
justified by the lack of a chemical bond between the carbon and
nitrogen atoms, this contribution is zero.  In our calculation we
neglect the contributions from fullerene electrons by considering
large fullerenes, and by choosing the cut-off radius $r_{c}$ used to
determine the integral in \eqnref{Nl}, to be sufficiently smaller than
the cage radius, $R$, but sufficiently large to enclose
the region one would chemically attribute to the endohedral atom.  We
discuss the choice of $r_{c}$, with particular attention to the
difference in applicability for N@C$_{60}$ and P@C$_{60}$ in the next
section.

Following the above discussion, the hyperfine compression ratio of
Buchachenko,\cite{Bucha01} originally given in terms of the endohedral
and free atom hyperfine coupling constants $a$ and $a_{0}$,
%BW punctuation
respectively, may then be expressed as
\bea\label{phi}
\phi & = & {{a-a_{0}} \over {a_{0}}} \\
& = & {{p_{l=0}^{unpaired}(endohedral) - p_{l=0}^{unpaired}(free)}
\over
{p_{l=0}^{unpaired}(free)}}
\eea

\noindent in which a consistent value of $r_{c}$ is used for the
evaluation of both the endohedral and free atom terms, in order to
make the probability density for the $l=0$ electron at the nucleus transferable
between the two cases, as discussed above.

\section{Results}\label{results}

\begin{figure}
\includegraphics[width = 8.5cm]{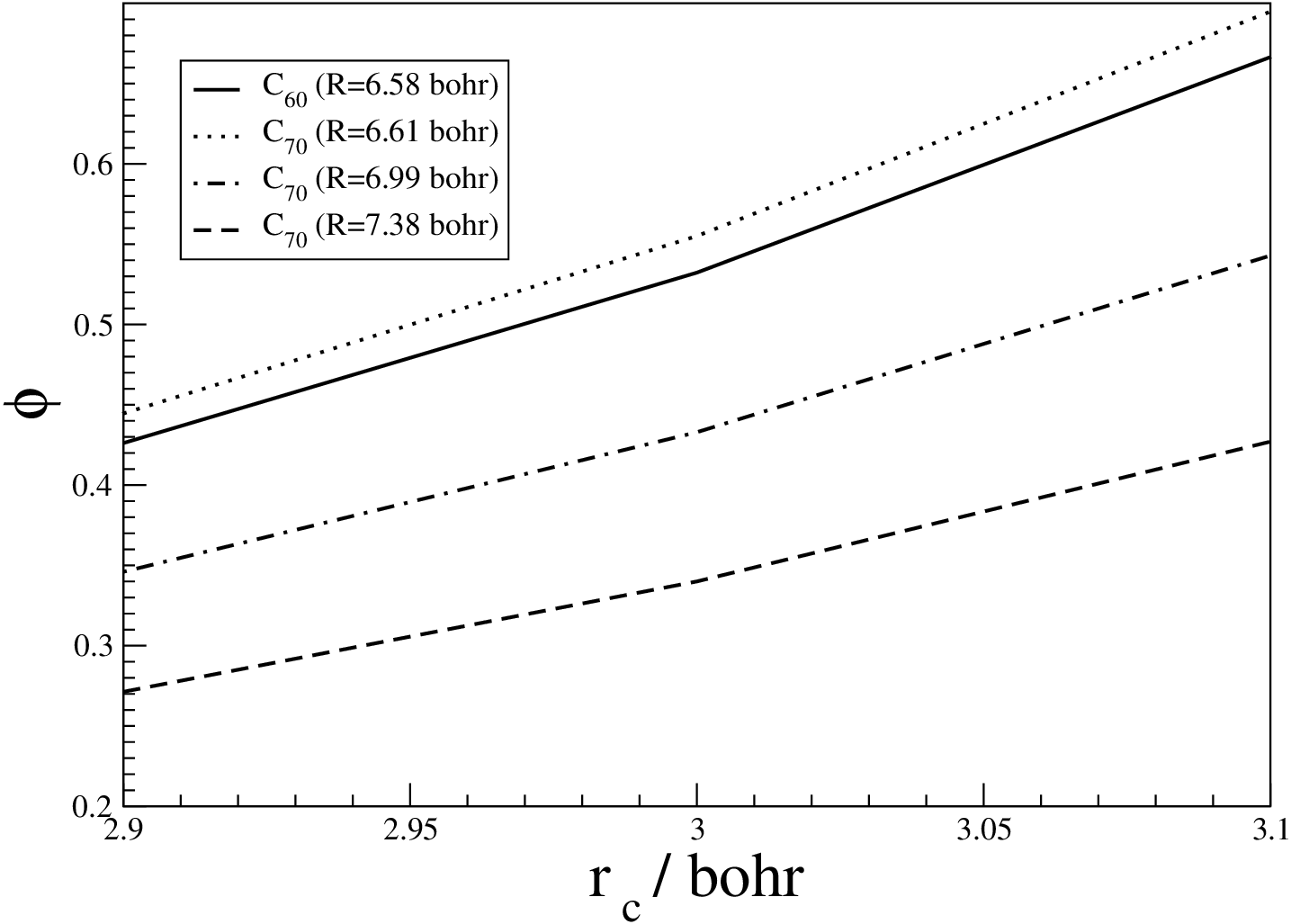}
\caption{
Hyperfine compression ratio, $\phi$, as a function of cutoff
radius, $r_c$.  For $r_c = 3.01$ bohr, the compression ratios calcualated for N@C$_{60}$
and the appropriate weighted average for N@C$_{70}$ are 0.541 and 0.490,
respectively, 
%BW
in excellent agreement with the experimental values 0.541 for N@C$_{60}$
and 0.491 for N@C$_{70}$.\cite{Bucha01}
} 
\label{rcdetermine}
\end{figure}

To determine the appropriate value of the cut-off radius, $r_{c}$, we
plot calculated values of $\phi$ versus $r_c$ for N@C$_{60}$ and
N@C$_{70}$ in \figref{rcdetermine}, obtained using the 
experimental fullerene
dimensions given by Buchachenko.\cite{Bucha01} To match the
experimental
measurement for N@C$_{60}$ of $\phi = 0.541$, we take $r_c = 3.01$
bohr. This yields values of $\phi = $ 0.541, 0.562 and 0.345 for
C$_{60}$, for 6.607 bohr radius (short axis) C$_{70}$, and for 7.38 bohr
radius (long axis) C$_{70}$, respectively.  Since the C$_{70}$
molecule is prolate, the average value of $\phi$ computed by weighting
the short axis value twice is 0.490, in excellent agreement with the
experimentally measured value of $\phi = 0.491$.\cite{Bucha01} For
comparison, we note that the values of $\phi$ 
evaluated with 
$r_{c} = 3.00$ bohr
are 0.531 and 0.486 for C$_{60}$ and C$_{70}$ (average value), and
also in reasonable agreement with experiment.  Our value of $r_{c}
\approx 3$ bohr is consistent with the spatial extent of the spin
density distribution determined in the {\it ab initio} calculations of
Kobayashi \etal\cite{KND03} However, our attempt to perform a similar
calculation for P@C$_{60}$ was unsuccessful, due to the larger size of
the P atom. This requires a larger $r_c$ and thus unavoidably
introduces possible contributions from fullerene electrons.  Thus we
restrict the analysis below to nitrogen endohedral fullerenes.

\begin{figure}
\includegraphics[width = 8.5cm]{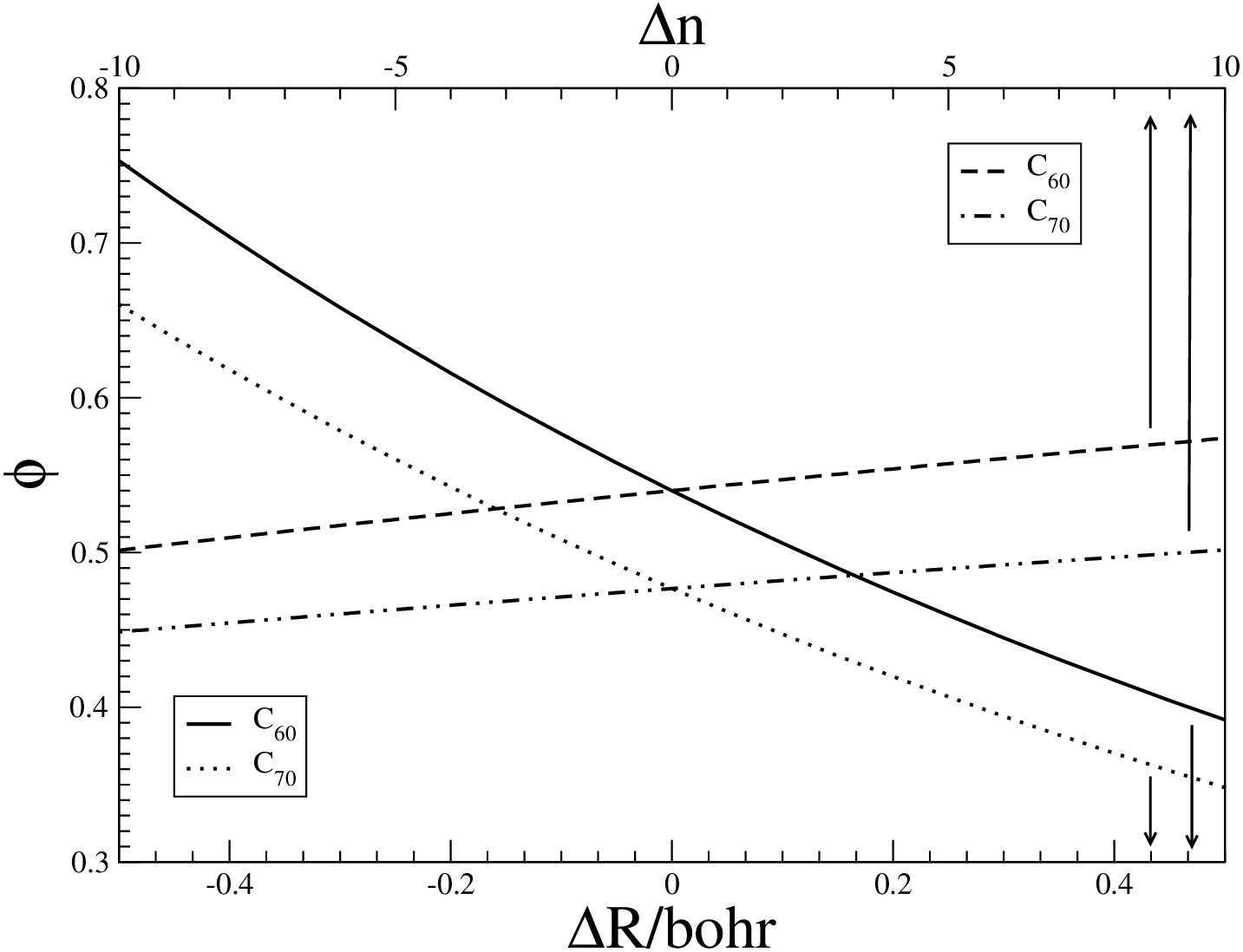}
\caption{Effects of changing $n$ (positive slope curves) and $R$
(negative slope curves) for C$_{60}$ ($n_0 = 60, R_0 = 6.578$ bohr) and a 
spherical model 
C$_{70}$ ($n_0 = 70, R_0 = 6.8647$ bohr) 
on the hyperfine compression ratio $\phi$.  
Here $n = n_0 + \Delta n$ and $R = R_0 + \Delta R$, where $n_0$, $R_0$ are the
equilibrium values of $n$ and $R$, respectively.
For C$_{60}$ at $\Delta n = 0$ 
and $\Delta R = 0$, $d\phi/dn = 0.0031$ and $d\phi/dR = -0.35$ bohr$^{-1}$.    }
\label{charge_vs_size}
\end{figure}

We have examined the relative roles of the number of cage
$\pi$-electrons, $n$, and radius, $R$, on $\phi$, shown in
\figref{charge_vs_size}.  Intuitively, one expects $\phi$ to be
inversely proportional to fullerene radius $R$ (at constant number of
cage electrons $n$), and to be directly proportional to $n$ (at
constant $R$).  Variations in $n$ and $R$ may arise experimentally by
endohedral inclusion into defect fullerenes,\cite{LH04} by use of
boron or nitrogen substitutions in place of carbon atoms in the
fullerene,\cite{MHOS92, CZS+98} or as the result of interactions with
surfaces.\cite{YM96,LGK+04} In practice, 
simultaneous variations of $n$ and $R$ are unavoidable. 
Nevertheless, we will analyze ideal
independent variations of these two parameters in order to establish their
relative importance.  The papers by Buchachenko\cite{Bucha01} and by
Kobayashi \etal,\cite{KND03} both make the qualitative statement that
cage radius ($R$) is more important than the cage electrons ($n$) in
enhancing the hyperfine coupling of the endohedral nitrogen atom.  The
simplicity of the present model allows us to separate and directly
test these two contributions.  \figref{charge_vs_size} confirms these
statements, as is visible from the much steeper slope of the
constant-$n$ curves (i.e., varying $R$), as compared to the
constant-$R$ (i.e., varying $n$) curves.  For N@C$_{60}$, at $\Delta n
= 0$ and $\Delta R = 0$, the slopes are $d\phi/dR = -0.35$ bohr$^{-1}$
and $d\phi/dn = 0.0031$.  We discuss the implications of the magnitude
of these dependencies for quantum computation in the next section.

\begin{figure}
\includegraphics[width = 8.5cm]{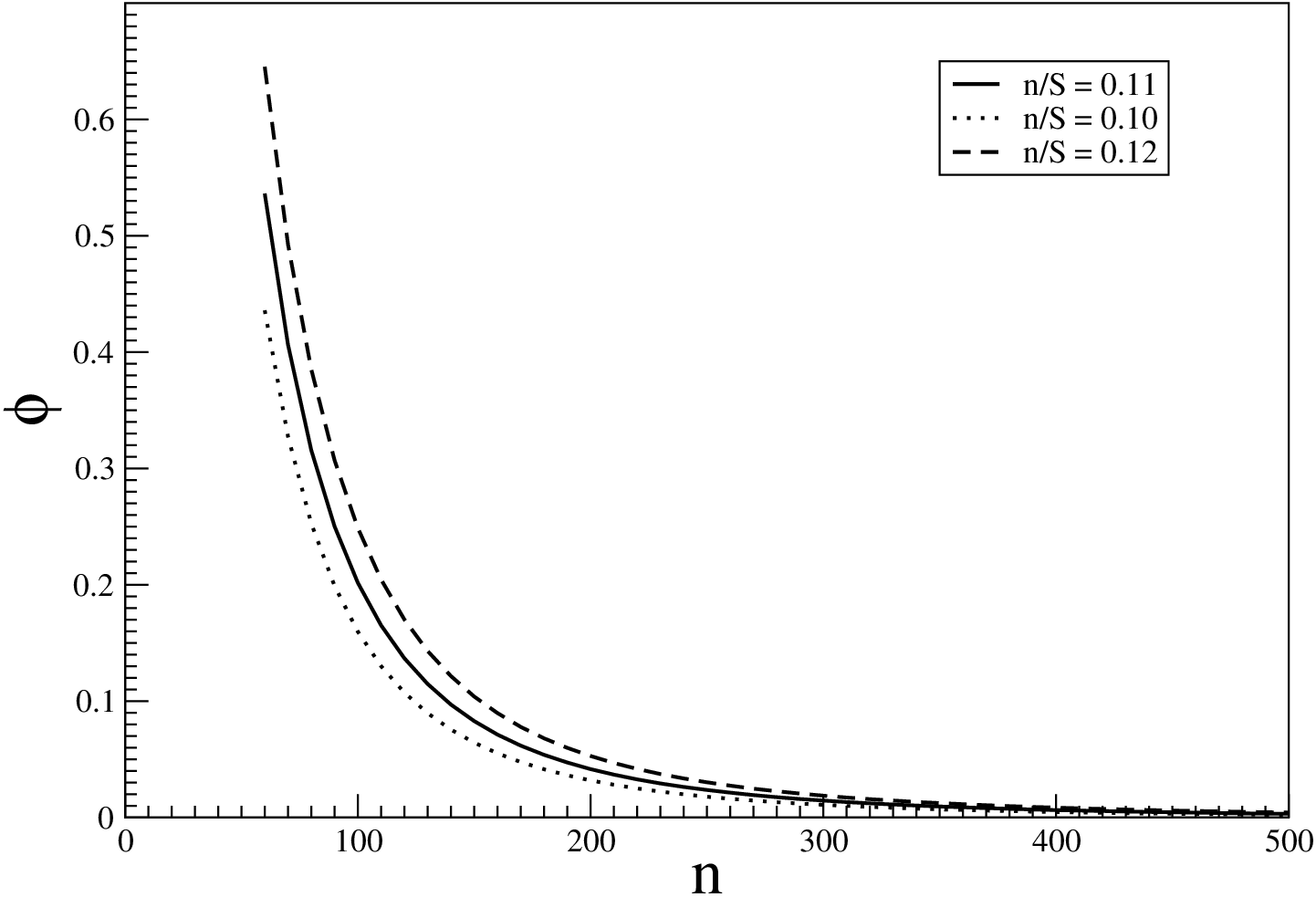}
\caption{Values of the hyperfine compression ratio, $\phi$ for the 
N@C$_{n}$ with $60 < n < 500$, assuming surface charge/area
ratios, $n/S$, of 0.10, 0.11, and 0.12.
}
\label{hypotheticalfullerenes}
\end{figure}

To examine the hypothetical endohedral fullerenes N@C$_{n}$ with $60 <
n < 500$, most of which have not yet been systematically studied, we
have assumed that the ratio of $n/S$ is constant, where $S = 4 \pi
R^{2}$ is the surface area of a spherical fullerene.  Thus for
C$_{60}$, $S = 543$ bohr$^{2}$ and $n/S = 0.11$; for C$_{70}$, $S =
592$ bohr$^{2}$ and $n/S = 0.12$.  \figref{hypotheticalfullerenes}
shows the behavior of the hyperfine compression ratio, $\phi$, for
N@C$_{n}$ with $60 < n < 500$, for $n/S = $ 0.10, 0.11, and 0.12.  As
expected, $\phi$ is seen to be a monotonically decreasing function of
$n$, regardless of the choice of $S/n$, and asymptotically approaches
zero (i.e., the free-atom limit of no enhancement of the hyperfine
coupling constant) as the fullerene becomes larger.  Depending on the
exact value of $S/n$, we find $\phi < 0.1$ for $130 \leq n \leq 160$
and $\phi < 0.01$ for $350 \leq n \leq 380$.  Even for the
hypothetical C$_{500}$ fullerene, with a radius of $\sim 40$ bohr,
%BW
we find a finite enhancement of
$0.002 \leq \phi \leq 0.004$, which could be detectable
experimentally.\cite{Bucha01}

\section{Summary and Discussion}\label{conclusion}

Using a modified Thomas-Fermi theory, we have developed a simple
physical model for the effect of endohedral inclusion on the hyperfine coupling constant of nitrogen.
The only free parameter in our model is the cutoff
radius, $r_{c}$, used to partition the atom from the fullerene. A
single choice of $r_c \approx 3.0$ bohr is found to give quantitative
agreement with 
%BW
recent experimental results for both N@C$_{60}$ and N@C$_{70}$.
Moreover, the simplicity of our model allows us to separate the
relative contributions of cage potential and size on the hyperfine
coupling constants, and to thereby identify the fullerene radius as the
dominating factor in the scaling of 
%BW
the hyperfine compression ratio
$\phi$.  We have also predicted the
hyperfine coupling constants for nitrogen endohedral to  
hypothetical fullerenes as large as C$_{500}$
%BW
and find that a finite enhancement should exist up to this size.

Spin exchange betwen the electron and nuclear degrees of freedom
mediated by the hyperfine interaction is necessary to implement the
quantum cellular automata scheme of Twamley.\cite{Twa03,FT04b} Since
global operations on all the qubits are used to evolve the
computational unitary in this scheme, deviations in the hyperfine
coupling constants of the individual sites, e.g. due to deformations
resulting from fullerene interaction with the substrate surface, would
require additional complication in the nuclear-electronic CNOT and
swap operations in order to be robust against these variations.  Our
results in \sectref{results} indicate that the hyperfine coupling is
quite sensitive to these deformations, and that even a 1\% (0.07\AA)
change in the N@C$_{60}$ radius would result in a $\sim$ 5\% change in
%BW
the compression ratio
$\phi$, corresponding to an approximately 0.5 MHz shift in the hyperfine 
constant.
Using the fidelity measure $F =\vert {\rm Tr}(VU^{\dagger})\vert /
{\rm Tr}(UU^{\dagger})$, where $U$ is the desired unitary and $V$ is
the actual (erroneous) unitary,\cite{MT04} a 0.5 MHz shift away from
the nominal hyperfine interaction gives $1-F = 2 \times 10^{-4}$ for
the electron-nuclear CNOT operation.  Atomistic calculations treating
the fullerene-surface interactions in more detail, may provide useful
estimates for the extent of this deformation on experimentally
relevant surfaces.

%Although current STM-based electron
%spin resonance methods for examining single molecules are not yet
%capable of hyperfine coupling determination,\cite{DW02,Durkan04}
%future experimental studies may be able to use the relationship shown
%in \figref{charge_vs_size} as a sensitive measure of cage size change
%due to surface binding.

\section{Acknowledgements}

J.S. thanks Professor Robert A. Harris for a helpful discussion, as well as 
the National Defense Science and
Engineering Grant (NDSEG) program and U.S. Army Research Office
Contract/Grant No. FDDAAD19-01-1-0612 for financial support.
K.B.W. thanks the Miller Institute for Basic Research in Science for
financial support.  This work was also supported by the Defense
Advanced Research Projects Agency (DARPA) and the Office of Naval
Research under Grant No. FDN00014-01-1-0826, and the National Science
Foundation under Grant EIA-020-1-0826.

%\bibliography{nc60,dft,betadecay}
%\bibliographystyle{apsrev}

\vfill\eject

\vspace{2cm}

\vspace{2cm}

\end{document}